Revolutions in science: The proposal of an approach for

the identification of most important researchers, institutions, and countries

based on Reference Publication Year Spectroscopy (RPYS)

Lutz Bornmann*$, Robin Haunschild$, Werner Marx$

* Science Policy and Strategy Department

Administrative Headquarters of the Max Planck Society

Hofgartenstr. 8,

80539 Munich, Germany.

Email: bornmann@gv.mpg.de

$ Max Planck Institute for Solid State Research

Heisenbergstraße 1,

70569 Stuttgart, Germany.

Email: l.bornmann@fkf.mpg.de, r.haunschild@fkf.mpg.de, w.marx@fkf.mpg.de

**Abstract**

RPYS is a bibliometric method originally introduced in order to reveal the historical roots of research topics or fields. RPYS does not identify the most highly cited papers of the publication set being studied (as is usually done by bibliometric analyses in research evaluation), but instead it indicates most frequently referenced publications – each within a specific reference publication year. In this study, we propose to use the method to identify important researchers, institutions and countries in the context of breakthrough research. To demonstrate our approach, we focus on research on physical modeling of Earth's climate and the prediction of global warming as an example. Klaus Hasselmann and Syukuro Manabe were both honored with the Nobel Prize in 2021 for their fundamental contributions to this research. Our results reveal that RPYS is able to identify most important researchers, institutions, and countries. For example, all the relevant authors' institutions are located in the USA. These institutions are either research centers of two US National Research Administrations (NASA and NOAA) or universities: the University of Arizona, Princeton University, the Massachusetts Institute of Technology (MIT), and the University of Stony Brook.

# 1      Introduction

Research is denoted as being breakthrough if it leads to discoveries with an enormous impact on future science (Winnink, Tijssen, & van Raan, 2018). Such discoveries are frequently connected to revolutionary changes of theory (Kuhn, 1962). Starting initially as claims of truth within frontier knowledge, discoveries usually become core knowledge later on that are characterized by universal consensus in a field (Cole, 1992; see also Collins, 1983). Research leading to breakthroughs is as a rule the result of creative solutions by researchers to complex problems (Hemlin, Allwood, Martin, & Mumford, 2013). The identification of research that can be called breakthrough by using bibliometric methods has been undertaken by Redner (2005), Schneider and Costas (2017), and Ponomarev, Williams, Hackett, Schnell, and Haak (2014). Studies that have investigated the impact of discoveries on technology development show that "it can take many years before a scientific discovery finds its way into new or adapted technology" (Winnink et al., 2018).

Although bibliometric methods have hitherto been used in some studies to identify breakthrough research, the standard approach to identification is qualitative: the reading of papers by experts (peers) in a certain field (Chubin & Hackett, 1990; Hammarfelt, Rushforth, & de Rijcke, 2020; Shibayama & Wang, 2020). The awarding of the Nobel Prize is a good example of using peers (in a peer review process) to identify groundbreaking research: Members of institutions awarding the prize meet to vote on potential nominees for the Nobel Prize fields. Only researchers with the most significant benefits to humankind in the fields relevant to the prize are in the position to receive the prize. Since some time is usually necessary to assess the benefits, the Nobel Prize is in most cases awarded many years after research has been done (Wagner, Whetsell, & Mukherjee, 2019).

Although scientific breakthroughs are usually well known to researchers in a field, it is frequently unclear which researchers, institutions and countries have actively participated in research that leads to breakthroughs. Core papers introducing breakthrough research might be well known in a field, but the contributing researchers, institutions, and countries are frequently not. In this study, therefore, we propose a bibliometric approach that can be applied for this purpose. The approach is based on a bibliometric instrument that has been developed by Marx, Bornmann, and Barth (2013) to identify the historical roots of research topics, research fields or research of eminent researchers: the so-called Reference Publication Year Spectroscopy (RPYS, see also Marx, 2021; Marx, Bornmann, Barth, & Leydesdorff, 2014). In this study, we use research on the physical modeling of Earth's climate and the prediction of global warming as an example in order to demonstrate our approach. Klaus Hasselmann and Syukuro Manabe were both honored with the Nobel Prize in 2021 for their fundamental contributions to this research.

## 2 Reference Publication Year Spectroscopy (RPYS)

Citation analysis is usually based on counting the number of citations a focal paper has received. Thus, it is a method for investigating citing papers published in the (near) future. RPYS instead focuses on cited references, i.e., publications that have been cited by focal papers in a set of selected publications. Since it is the objective of RPYS to explore the historical roots of research topics or fields, the compilation of the publication set representing certain topics or fields is the first step in RPYS. This is the most important step in RPYS: Historical roots can only be reliably and validly measured if papers explicitly dealing with a topic or field are considered. The consideration of too many peripheral publications in the publication set would lead to mistakes and a lack of clarity when identifying historical roots. In the second step of RPYS, the cited references of the publications in the set are analyzed. Publications that appeared decades ago and which have been cited many times from the publications in the field- or topic-specific set of publications can be denoted as historical roots of the field or topic.

RPYS can be undertaken by using the Cited References Explorer program (CRExplorer, www.crexplorer.net). Since the introduction of RPYS by Marx et al. (2013) and Marx et al. (2014), the method has been frequently applied in different contexts: Historical roots have been identified in women's studies (Yun, Lee, & Ahn, 2020), innovation in creative industries (Gohoungodji & Amara, 2022), iMetrics (bibliometrics) in China (Li, Yao, Tang, Li, & Wu, 2020), the concept of health equity (Yao et al., 2019), critical social psychology in Brazil (Millán et al., 2020), social network analysis (Camacho, Panizo-LLedot, Bello-Orgaz, Gonzalez-Pardo, & Cambria, 2020), and modern plastic surgery (Chopan, Sayadi, Buchanan, Katz, & Mast, 2019). Historical roots of a journal have been investigated by Haunschild (2019), and the roots of an eminent single researcher by Leydesdorff, Bornmann, Comins, Marx, and Thor (2016). These studies are only a selected set of studies that have applied RPYS in various fields and topics. In recent years, the application of RPYS was very popular in general, due – we assume – to the simplicity of its use in a well-defined research field or topic.

Some studies have investigated whether the results of RPYS correspond to assessments by peers: Do peers identify the same publications as RPYS? Congruent results have been reported by Vosner, Carter-Templeton, Zavrsnik, and Kokol (2020) for the field of nursing informatics, Yeung (2020) for carotid artery stenting which is a treatment option for carotid artery stenosis, and by Kokol, Blažun Vošner, and Završnik (2021) for the application of bibliometrics in medicine. The latter authors concluded, e.g., that "RPYS proved to be an accurate method, which was able to identify most of the recognised historical roots of the application of bibliometrics in medicine". The results of these studies are especially important for the application of RPYS, since they demonstrate that RPYS leads to valid results. Thus, the results indicate that the method can be used to identify historical roots in research fields and topics.

# 3 Methods

## 3.1 Reference Publication Year Spectroscopy based on co-citations (RPYS-CO)

In this paper, we used RPYS-CO which is a specific form of RPYS and which is based on a restricted set of cited references: those which are co-cited with one or more specific papers. Thus, the method is based on the co-citation network of publications (Small, 1973). The method takes advantage of the fact that concurrently cited (co-cited) papers are more or less closely related to each other. One can select the citation environment of a specific paper in the form of all co-cited papers and analyze these papers as cited references applying RPYS-CO. The specific papers should be prominent seminal works, which can be used as marker or tracer papers for a specific topic or field (in this study: research on physical modeling of Earth's climate and the prediction of global warming). We assume that papers co-cited with the selected papers are potential candidates for being relevant in a specific historical context.

RPYS-CO was first introduced in connection with the history of the greenhouse effect (Marx, Haunschild, Thor, & Bornmann, 2017) and was later successfully tested in theoretical chemistry (density functional theory, see Haunschild & Marx, 2020) and climate change and tea production (Marx, Haunschild, & Bornmann, 2017). As a first step of RPYS-CO, the citing papers of a specific marker paper are downloaded from an appropriate literature database and all references cited therein are selected. The reference set contains all papers in the form of their references, which have been co-cited with the marker paper. In cases where there is more than one marker paper, the papers of the publication set are co-cited with at least one of the marker papers.

RPYS and RPYS-CO usually focus on single publications that have been frequently cited in a field- or topic-specific publication set: One is interested in the identification of core publications or historical roots. In this study, the use of the RPYS-CO approach has another focus: It is not single publications but the complete ensemble of publications cited. We interpret the complete ensemble as the contributing publications to a certain breakthrough in a field: Without the various research from the contributing publications, the breakthrough would not have been possible. Since contributing publications are assigned to affiliations, affiliations can be analyzed with respect to researchers, institutions, and countries. Frequently emerging researchers, institutions, and countries can then be interpreted as the most active and contributing units to a certain breakthrough.

We plotted the RPYS spectrogram that is presented in this study by using R (R Core Team, 2018) with the R package 'BibPlots' (Haunschild, 2021). In addition to the static plot in this paper, we produced an interactive RPYS graph (Haunschild & Bornmann, 2021) using the R package 'dygraphs' (Vanderkam, Allaire, Owen, Gromer, & Thieurmel, 2018).

## 3.2 Research on the physical modeling of Earth's climate and the prediction of global warming

Since the availability of powerful computers from around 1970, climate modeling has become a rapidly increasing branch of climate change research. Quantitative climate models simulate the interactions of major climate system components (i.e., atmosphere, land surface, ocean, and sea ice) taking account of the incoming energy from the sun. Scientists use climate models to understand the complex climate system of the Earth, to test their hypotheses, and to draw conclusions on past and future climate. Climate models vary in complexity, ranging from simple radiant heat transfer models to coupled atmosphere-ocean-sea ice global climate models (GCMs). Predictions of climate models have become most important for strategies regarding adaption and mitigation of global warming.

Klaus Hasselmann and Syukuro Manabe were awarded jointly the Nobel Prize in Physics 2021: The prize was awarded "for groundbreaking contributions to our understanding of complex systems" (NobelPrize.org, 2021) with one half jointly to Syukuro Manabe and Klaus Hasselmann "for the physical modelling of Earth's climate, quantifying variability and reliably predicting global warming" (NobelPrize.org, 2021) and the other half to Giorgio Parisi "for the discovery of the interplay of disorder and fluctuations in physical systems from atomic to planetary scales" (NobelPrize.org, 2021).

**3.3     Dataset used**

For the RPYS-CO analysis in this study, we used seven seminal papers with the WoS document type "article" which were published by Klaus Hasselmann and Syukuro Manabe as marker papers. We selected these papers as follows: We retrieved papers by Klaus Hasselmann (K* Hasselmann) und Syukuro Manabe (S* Manabe) published between 1960-1980 by author searching in the Web of Science (WoS) and ranked them by citation counts. To focus on most relevant papers from the authors, we checked the papers with more than 300 citations for their relevance to climate modeling in connection with global warming. We selected six relevant papers for this study. Because of its great relevance, we additionally included the paper by Cubasch et al. (1992) that deals with greenhouse warming computations in our set of marker papers (see Table 1).

Table 1. Seven marker papers for the RPYS-CO analysis

| No. | Marker paper |
|---|---|
| 1 | Cubasch, U., Hasselmann, K. et al. (1992). Time-dependent greenhouse warming comutations with a coupled ocean-atmosphere model. *Climate Dynamics, 8*(2), 55-69. DOI: 10.1007/BF00209163 |
| 2 | Frankignoul, C. & Hasselmann, K. (1977). Climate models 2, application to sea-surface temperature anomalies and thermocline veriability. *Tellus, 29*(4), 289-305. DOI: 10.3402/tellusa.v29i4.11362 |
| 3 | Hasselmann, K. (1976). Stochastic climate models 1, theory. *Tellus, 28*(6), 473-485. DOI: 10.1111/j.2153-3490.1976.tb00696.x |
| 4 | Manabe, S. & Stouffer, R. J. (1980). Sensitivity of a global climate model to an increase of $CO^2$ concentrations in the atmosphere. *Journal of Geophysical Research – Oceans, 85*(NC10), 5529-5554. DOI: 10.1029/JC085iC10p05529 |
| 5 | Manabe, S. & Wetherald, R. T. (1967). Thermal equilibrium of atmosphere with a given distribution of relative humidity. *Journal of Atmospheric Sciences, 24*(3), 241-259. DOI: 10.1175/1520-0469(1967)024<0241:TEOTAW>2.0.CO;2 |
| 6 | Manabe, S. & Wetherald, R. T. (1975). Effects of doubling $CO^2$ concentration on climate of a general circulation model. *Journal of the Atmospheric Sciences, 32*(1), 3-15. DOI: 10.1175/1520-0469(1975)032<0003:TEODTC>2.0.CO;2 |

| 7 | Manabe, S. & Wetherald, R. T. (1980). Distribution of climate change resulting from an increase in $CO^2$ content of the atmosphere. *Journal of the Atmospheric Sciences, 37*(1), 99-118. DOI: 10.1175/1520-0469(1980)037<0099:OTDOCC>2.0.CO;2 |
|---|---|

The seven marker papers in Table 1 have been cited by 4,340 publications that are indexed in the WoS (date of search: November 02, 2021). We downloaded these citing publications from the WoS and imported them into CRExplorer. We extracted 233,586 non-distinct cited references from the 4,340 publications. The number of distinct cited references is 115,377. We removed all cited references that were published after 1980 to focus on the time of the invention under study and earlier time periods. We used the automatic clustering and merging using CRExplorer with a Levenshtein threshold of 0.75. We used volume and (starting) page number in the clustering and merging process to have an exact match. To sharpen the spectrogram, we removed all cited references that were cited less than five times. We additionally removed the marker papers so that they could not distort the results. We finally retained a set of 1,424 cited references that we used for the analyses in this study.

## 4 Results

In the next two subsections, we present two different views on the RPYS-CO results: subsection 4.1 presents a quantitative and subsection 4.2 a qualitative view on important researchers, institutions, and countries in the context of research on physical modeling of Earth's climate and the prediction of global warming.

### 4.1 Important researchers, institutions, and countries from our RPYS results

We present the results of three quantitative analyses in this subsection. Firstly, we analyzed important researchers, institutions, and countries using our full RPYS-CO dataset (i.e., all cited references that were referenced at least five times). Secondly, we restricted the analysis of these entities to the top-10 cited references of each reference publication year.

Thirdly, we restricted the analysis to the top-5 cited references of each reference publication year. In the following, we show only the top entities from the analyses. The full lists are available as Supplemental Information (see https://s.gwdg.de/q7jUGm).

For all three analyses, we extracted the DOIs from the cited references and searched the DOIs in the WoS. We used the built-in analyze function of the WoS for obtaining the statistics. Unfortunately, not all cited references had a DOI. Considering that the DOIs were assigned in retrospect, surprisingly many cited references have DOIs. Out of all 1,424 cited references published between 1686 and 1980, 72.1% (n=1,026) of the cited references had a DOI. However, not all DOIs could be found in the WoS. More importantly, not all retrieved cited references in the WoS had affiliation information attached to them. For our dataset including all 1,424 cited references, where 1,026 have a DOI, 922 of those DOIs could be found in the WoS, but 29.4% (n=271) of them do not have affiliation information in the WoS. This affected the country and institutional analyses, since information on country and institution was available for only around 70% of the cited references that could be found in the WoS. However, the analyses on the author-level were not affected. In the case of the top-10 cited references from each reference publication year, 200 DOIs were found in the WoS but 63.5% (n=127) of them did not have affiliation information attached to them. In the case of the top-5 cited references, the proportion of cited references without affiliation information in the WoS is slightly higher with 64.1% (75 out of 117 cited references did not contain affiliation information).

Table **2** shows the most important authors, institutions, and countries of authors using all cited references in our dataset. Note that "Fed. Rep. Ger.", "West Germany", and "Germany" were merged into the country name "Germany". Table 3 shows the same set of information after restricting our dataset to the top-10 cited references of each reference publication year, and Table 4 shows the corresponding analyses when using only the top-5 cited references of each reference publication year.

Table 2. Most frequently co-cited authors, institutions, and countries using all cited references in our dataset (NCR = number of cited references)

| Author | NCR | Institution | NCR | Country | NCR |
|---|---|---|---|---|---|
| Manabe, S. | 25 | National Center Atmospheric Research (NCAR, USA) | 103 | USA | 531 |
| Schneider, S. H. | 19 | National Oceanic Atmospheric Administration (NOAA, USA) | 59 | England | 42 |
| Ramanathan, V. | 15 | National Aeronautics Space Administration (NASA, USA) | 43 | Australia | 32 |
| Crutzen, P. J. | 14 | University of California System (USA) | 42 | Germany | 20 |
| Namias, J. | 14 | Massachusetts Institute of Technology (MIT, USA) | 35 | Canada | 10 |

Table 3. Most important authors, institutions, and countries using the top-10 cited references of each reference publication year (NCR = number of cited references)

| Author | NCR | Institution | NCR | Country | NCR |
|---|---|---|---|---|---|
| Manabe, S. | 11 | National Center Atmospheric Research (NCAR, USA) | 16 | USA | 64 |
| Ramanathan, V. | 8 | National Aeronautics Space Administration (NASA, USA) | 9 | England | 3 |
| Schneider, S. H. | 8 | National Oceanic Atmospheric Administration (NOAA, USA) | 8 | Switzerland | 3 |
| Keeling, C. D. | 6 | Princeton University (USA) | 8 | | |
| Lorenz, E. N. | 5 | | | | |

Table 4. Most important authors, institutions, and countries using the top-5 cited references of each reference publication year (NCR = number of cited references)

| Author | NCR | Institution | NCR | Country | NCR |
|---|---|---|---|---|---|
| Manabe, S. | 7 | National Center Atmospheric Research (NCAR, USA) | 9 | USA | 36 |
| Ramanathan, V. | 7 | National Aeronautics Space Administration (NASA, USA) | 5 | England | 2 |
| Schneider, S. H. | 5 | Princeton University (USA) | 5 | Switzerland | 2 |
| Cess, R. D. | 4 | | | | |

Table **2** to Table **4** reveal three authors most frequently co-cited with the marker papers, which appear in each of the tables: Syukuro Manabe, Veerabhadran Ramanathan, and Stephen H. Schneider. Explanations to Syukuro Manabe as one of the physics Nobel laureates in 2021 has been given above. The results in the tables reveal that some of his papers are strongly related to the selected marker papers and are frequently co-cited with them. Klaus Hasselmann, however, does not appear among the top co-cited references of this analysis. Other relevant papers do not seem to exist besides the marker papers.

Veerabhadran Ramanathan (1944-) at Scripps Institution of Oceanography (University of California in San Diego) and member of the US National Academy of Sciences has contributed to many areas of atmospheric and climate sciences, including the development of general circulation models and the detection of climate change. In 1975, he discovered the strong greenhouse effect of cholorofluorocarbons (CFCs). Before this discovery, carbon dioxide was widely seen to be the only greenhouse gas causing global warming. Stephen H. Schneider (1945-2010, since 1992 at Stanford University and member of the US National Academy of Sciences) made fundamental contributions to the understanding of human effects on Earth's climate. He showed that human beings are active agents of climate change. He was one of the first to appreciate the importance and power of climate modeling. One of his research topics relates to the statistical properties of climate model simulations, seeking to separate anthropogenic climate change signals from the noise of natural climate variability – the so-called signal to noise problem (National Academy of Sciences, 2014).

Robert D. Cess published on atmospheric feedback mechanisms and also appeared as a co-author of Veerabhadran Ramanathan. The appearance of Charles David Keeling (originator of the Keeling curve which shows the increase of the carbon dioxide concentration in the atmosphere), of Paul Josef Crutzen (chemistry Nobel laureate in 1995 for his work on atmospheric chemistry, particularly on the decomposition of atmospheric ozone), and Edward

Norton Lorenz (discoverer of the butterfly effect, i.e., the sensitive dependence of complex systems such as climate on initial conditions) indicate the broad perspective when analyzing all co-cited references or all top co-cited references of each reference publication year. The contributions of these scientists are so fundamental that they can be expected to appear in bibliometric analyses of the history of climate change research.

In the 1970s and early 1980s, the vast potential of numerical models of the climate system was becoming increasingly obvious. According to Moore's law, computing power has increased tremendously. Moore's law is an empirical relationship, predicting that the number of components per integrated circuit doubles about every two years, developing in line with a corresponding increase of computing power and storage capacity. During the 1970s, however, only very few research centers were able to perform such modeling. Therefore, the top-ranked institutions are a small number of large national research centers, all of which are located in the USA (NCAR, NOAA, NASA). The famous Hadley Centre in England did not appear, because it was not founded until 1990 (it was initially named the Hadley Centre for Climate Research and Prediction, and was later renamed). There is a long tradition of climate research not only in England, but also in Switzerland. Therefore, besides the USA, England and Switzerland also appeared among the top-ranked countries of authors of the co-cited references.

The statistical analyses based on all co-cited references or all top co-cited references of each reference publication year, respectively, detect a multitude of papers that have strong and weak connections to climate modeling. The latter are historical or fundamental works of climate research that do not focus on climate simulations. In order to focus more specifically on the immediate intellectual environment of the beginning of climate modeling, we have additionally analyzed distinct peaks in the RPYS-CO spectrogram. The peaks that we interpret in the next section reveal the most important publications that are directly related to the seven marker papers.

## 4.2 Qualitative analysis of the RPYS-CO

We applied RPYS-CO and retrieved the spectrogram of all co-citations of the seven marker papers (see Figure 1).

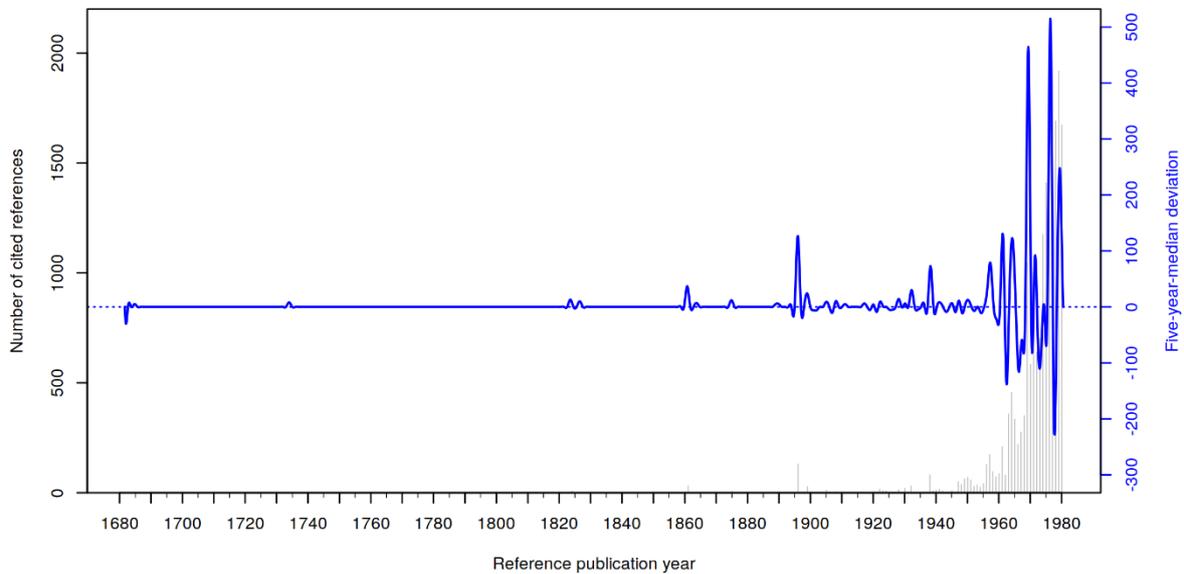

Figure 1. RPYS-CO spectrogram of the seven marker papers listed in Table 1. The marker papers are not included in this spectrogram (online figure at https://s.gwdg.de/tflsR5)

The most-frequently co-cited publications, which appear as distinct peaks in the RPYS-CO spectrogram of Figure 1, can be broadly categorized as follows:

1. Historical works relating to the discovery of the greenhouse effect which are most important for the history of global warming (Arrhenius, 1896; Chamberlin, 1899; Croll, 1864, 1875; Fourier, 1824, 1827; Hadley, 1735; Tyndall, 1861).

2. Papers published since the 1930s and dealing with the carbon dioxide theory of climate change (Callendar, 1938; Plass, 1956; Revelle & Suess, 1957) and effects of various parameters (e.g.. ocean currents, solar cycles) on the mean temperature of the atmosphere (Manabe, 1969; Stommel, 1961; Walker & Bliss, 1932). Budyko (1969) is not directly related to global warming. Edward Lorenz is the discoverer of the butterfly effect. The work by Lorenz (1963) is fundamental for the prediction of future weather and climate (without a distinct peak in the spectrogram).

3. Papers published between 1960 and 1980 deal with climate models and calculations of the mean Earth temperature as a function of the concentration of greenhouse gases, particularly of carbon dioxide. The papers focus on climate models (Sellers, 1969), aerosols (Rasool & Schneider, 1971), ice ages, feedback mechanisms, man-made greenhouse gases, and radiative convection (Cess, 1976; Hays, Imbrie, & Shackleton, 1976; Ramanathan, 1976; Wang, Yung, Lacis, Mo, & Hansen, 1976), as well as models of the influence of $CO_2$ on the atmospheric temperature (Manabe & Stouffer, 1979; Newell & Dopplick, 1979; Ramanathan, Lian, & Cess, 1979).

Most papers published between 1960 and 1980 belong to the narrow periphery of the marker papers published by Klaus Hasselmann and Syukuro Manabe and are closely related to their works. They can be interpreted as the intellectual environment of the beginning of climate modeling. Besides Klaus Hasselmann and Syukuro Manabe, in particular Veerabhadran Ramanathan, Richard Wetherald, and William Sellers can be seen as further pioneers of climate modeling (only Veerabhadran Ramanthan is still alive).

The contributions of Veerabhadran Ramanathan to climate modeling have been mentioned above. Richard Wetherald (1936-2011) appears as a co-author of Syukuro Manabe on three of the seven marker papers that we used in this study. He was an American climate scientist who made fundamental contributions to climate modeling. William D. Sellers (1928-2014) was an American meteorologist and climate scientist who created one of the first climate models. He was one of the first who discussed the effect of atmospheric carbon dioxide on Earth's climate. Sellers (1969) paper entitled "A global climatic model based on the energy balance of the earth-atmosphere system" exactly hits the topic of climate modeling. In this paper, a relatively simple numerical model of the energy balance of Earth's atmosphere is applied. Besides Budyko (1969), Sellers (1969) caused the distinct 1969 peak in the spectrogram (see Figure 1).

Rasool and Schneider (1971) discuss the effects of large increases of atmospheric carbon dioxide and aerosols on the global climate. The papers by Hays et al. (1976) and Cess (1976) deal with the causes of ice ages and with atmospheric feedback mechanisms, respectively. With theses topics, the papers have a weaker connection with climate modeling. Wang et al. (1976) deal with the man-made greenhouse effect. The study used a one-dimensional radiative-convective model to simulate the effect of increased atmospheric trace gas concentrations on the Earth's surface temperature. Ramanathan (1976) presents a simplified radiative-convection model of the radiative transfer in the Earth's atmosphere.

Ramanathan et al. (1979) simulate zonal and seasonal effects on surface temperature caused by increased atmospheric carbon dioxide, based on radiative transfer model calculations. Manabe and Stouffer (1979) present a sensitivity study of the effects of carbon dioxide on Earth's climate based on a global climate model. The paper hits the topic of climate modeling but it was not used as a marker paper in this study: it received relatively low citation counts (93 citations until March 2022). Newell and Dopplick (1979) discuss "questions concerning the possible influence of anthropogenic $CO_2$ on atmospheric temperature". The authors simulated estimates of the atmospheric temperature changes due to a doubling of carbon dioxide concentrations with a static radiative flux model.

The results of the qualitative analyses in this section show that the intellectual periphery of research by Klaus Hasselmann and Syukuro Manabe is very narrow. All research-focussed institutions of the top-co-cited papers in our dataset are located in the USA: the Goddard Space Flight Center (NY) and Langley Research Center (Hampton, VA) of the National Aeronautics and Space Administration (NASA), and the National Center for Atmospheric Research of the National Oceanic and Atmospheric Administration (NOAA) in Boulder (CO). The two US National Research Administrations (NASA and NOAA) have a long-lasting tradition with regard to research related to climate change. In addition to these US institutions, the following well-known universities are most important: the University of Arizona (Tucson, CO), Princeton University (NJ), the Massachusetts Institute of Technology (MIT) in Cambridge (MA) in cooperation with the Scott Air Force Base (IL), and the University of Stony Brook (NY).

## 5 Discussion

RPYS is a bibliometric method originally introduced in order to reveal the historical roots of research topics or fields (Marx et al., 2013; Marx et al., 2014). The method aims to mirror the knowledge base of a specific field by analyzing the cited references within the papers of the relevant publication set (rather than the citations of the papers themselves). RPYS does not identify the most highly cited papers of the publication set under study (as usually done by bibliometric analyses in research evaluation), but rather the most frequently referenced publications – each in its specific reference publication year. The most frequently cited references are presented in graphical and tabular forms. The RPYS method provides an (objective) answer to questions concerning seminal papers and historical roots and is based on Galton's (1907) principle "wisdom of the crowd" (see Bornmann & Marx, 2014). Individual scientists in the field can answer such questions only subjectively, but the entire scientific community with knowledge in the target field can deliver a reliable view.

Whereas RPYS analyses have mainly focused on single seminal papers and the historical roots of research fields, we followed a proposal by Zhao, Han, Du, and Wu (2020) and others to extend classical RPYS. In this study, we used the method for identifying associated papers with revolutionary science or breakthrough research; we were specifically interested in developing and testing a method for identifying researchers, institutions, and countries involved in revolutionary science. This science usually does not occur "out of the blue" but is (deeply) embedded in previous and current research activities that are reflected in publications (Holton, Chang, & Jurkowitz, 1996). According to Yan, Tian, and Zhang (2020) "the source of novelty is often reflected in the new combination of existing knowledge and generation of new ideas" (p. 896). This view of novelty (or breakthrough) as a recombination of existing knowledge is in the tradition of Schumpeter (1939): Research processes result "in novelty when knowledge is combined 'differently' or when 'new combinations' are carried out" (Fontana, Iori, Montobbio, & Sinatra, 2020). Recombined novel research is in the unique position to push knowledge frontiers (Fontana et al., 2020) which are research topics of particular interest to researchers in a field (Fang, Costas, Tian, Wang, & Wouters, 2020).

In this study, we used research on the physical modeling of Earth's climate and the prediction of global warming as an example to demonstrate our new RPYS approach. We identified associated papers of research by Klaus Hasselmann and Syukuro Manabe who were honored with the Nobel Prize for their fundamental contributions to this research. Frequently co-cited references of research by Klaus Hasselmann and Syukuro Manabe (as well as associated papers) can be seen as seminal publications for research on the physical modeling of Earth's climate and the prediction of global warming.

The results of the quantitative reference analyses (i.e., of all co-cited references of the marker papers or only of the top co-cited references) as well as of the qualitative reference analysis (i.e., the co-cited peak references in the RPYS-CO spectrogram) show that the close intellectual environment of research by Klaus Hasselmann and Syukuro Manabe is rather narrow. Veerabhadran Ramanathan, Richard Wetherald, and William Sellers can be seen as further pioneers of climate modeling. Other notable researchers are: Robert D. Cess, Thomas G. Dopplick, James E. Hansen, Andrew A. Lacis, Mengshui. Lian, Tengfei Mo, Reginald E. Newell, S. Ichtiaque Rasool, Stephen H. Schneider, Ronald J. Stouffer, Wei-Chyung Wang, and Yuk L. Yung (in alphabetic order). All authors' institutions of the top-co-cited papers are located in the USA. These institutions are either research centers of two US National Research Administrations (NASA and NOAA) or universities: the University of Arizona,

Princeton University, the Massachusetts Institute of Technology (MIT), and the University of Stony Brook.

This study has two main limitations: (1) the choice of marker papers determines the set of cited references that are analyzed. One could choose to use more or less or different marker papers than we did. This might lead to different results. (2) The analyses of countries, institutions, and authors of the cited references can be done on different parts of the dataset. Either one uses all cited references from the RPYS analyses or a certain subset (e.g., top-5 or top-10 cited references from each RPY). A smaller subset might be too restrictive and a larger subset (or even the full set of cited references) might dilute the resulting analyses. Therefore, we performed quantitative analyses using three different datasets and an additional qualitative analysis.

The focus of this study was not on investigating the historical roots or the environment of climate change and global warming research. We used this research as an example to demonstrate our proposed method that is connected to RPYS. In principle, the method can be used for the investigation of any (revolutionary) research topic in any discipline. Since the Nobel Prize is usually awarded for breakthrough research, the method can be used to investigate important researchers, institutions, and countries in the context of rewarded research. As long as the publications of this research are covered in citation indexes, the proposed method can be applied.

# Acknowledgements

Lutz Bornmann is member of the Price Medal Laureate Board of *Scientometrics*, and Robin Haunschild is member of the Editorial Board of *Scientometrics*.

# References

Arrhenius, S. (1896). On the influence of carbonic acid in the air upon the temperature of the ground. *Philosophical Magazine and Journal of Science Series, 5*(41), 237-276.

Bornmann, L., & Marx, W. (2014). The wisdom of citing scientists. *Journal of the Association For Information Science and Technology, 65*(6), 1288-1292. doi: 10.1002/asi.23100.

Budyko, M. I. (1969). Effect of solar radiation variations on climate of earth. *Tellus, 21*(5), 611-619. doi: 10.3402/tellusa.v21i5.10109.

Callendar, G. S. (1938). The artificial production of carbon dioxide and its influence on temperature. *Quarterly Journal of the Royal Meteorological Society, 64*, 223-237.

Camacho, D., Panizo-LLedot, A., Bello-Orgaz, G., Gonzalez-Pardo, A., & Cambria, E. (2020). The four dimensions of social network analysis: An overview of research methods, applications, and software tools. *Information Fusion, 63*, 88-120.

Cess, R. D. (1976). Climate change: Appraisal of atmospheric feedback mechanisms employing zonal climatology. *Journal of the Atmospheric Sciences, 33*(10), 1831-1843. doi: 10.1175/1520-0469(1976)033<1831:Ccaaoa>2.0.Co;2.

Chamberlin, T. C. (1899). An attempt to frame a working hypothesis on the cause of glacial periods on an atmospheric basis. *Journal of Geology, 7*, 545-584.

Chopan, M., Sayadi, L., Buchanan, P. J., Katz, A. J., & Mast, B. A. (2019). Historical roots of modern plastic surgery: A cited reference analysis. *Annals of Plastic Surgery, 82*(6S), S421-S426. doi: 10.1097/sap.0000000000001666.

Chubin, D., & Hackett, E. (1990). *Peerless science: Peer review and U.S. science policy*. Albany, NY, USA: State University of New York Press.

Cole, S. (1992). *Making science. Between nature and society*. Cambridge, MA, USA: Harvard University Press.

Collins, H. M. (1983). The sociology of scientific knowledge: Studies of contemporary science. *Annual Review of Sociology, 9*, 265-285.

Croll, J. G. A. (1864). On the physical cause of the change of climate during geological epochs. *The London, Edinburgh, and Dublin Philosophical Magazine and Journal of Science, 28*(187), 121-137.

Croll, J. G. A. (1875). *Climate and time, in their geological relations, a theory of secular changes of the earth's climate*. London, UK: Edward Stanford.

Cubasch, U., Hasselmann, K., Höck, H., Maier-Reimer, E., Mikolajewicz, U., Santer, B. D., & Sausen, R. (1992). Time-dependent greenhouse warming computations with a coupled ocean-atmosphere model. *Climate Dynamics, 8*(2), 55-69. doi: 10.1007/BF00209163.

Fang, Z., Costas, R., Tian, W., Wang, X., & Wouters, P. (2020). An extensive analysis of the presence of altmetric data for Web of Science publications across subject fields and research topics. *Scientometrics, 124*(3), 2519-2549. doi: 10.1007/s11192-020-03564-9.

Fontana, M., Iori, M., Montobbio, F., & Sinatra, R. (2020). New and atypical combinations: An assessment of novelty and interdisciplinarity. *Research Policy, 49*(7), 104063. doi: 10.1016/j.respol.2020.104063.

Fourier, J. B. J. (1824). Remarques générales sur les températures du globe terrestre et des espaces planétaires. *Annales de Chimie et de Physique, 27*, 136–167.

Fourier, J. B. J. (1827). Mémoire sur les températures du globe terrestre et des espaces planétaires. *Mémoires de l'Académie Royale des Sciences, 7*, 569–604.

Frankignoul, C., & Hasselmann, K. (1977). Stochastic climate models 2, application to sea-surface temperature anomalies and thermocline variability. *Tellus, 29*(4), 289-305. doi: 10.1111/j.2153-3490.1977.tb00740.x.

Galton, F. (1907). Vox populi. *Nature, 75*, 450-451. doi: 10.1038/075450a0.

Gohoungodji, P., & Amara, N. (2022). Historical roots and influential publications in the area of innovation in the creative industries: A cited-references analysis using the Reference Publication Year Spectroscopy. *Applied Economics, 54*(12), 1415-1431. doi: 10.1080/00036846.2021.1976388.

Hadley, G. (1735). VI. Concerning the cause of the general trade-winds. *Philosophical Transactions of the Royal Society of London A: Mathematical, Physical and Engineering Sciences, 39*(437), 58-62. doi: 10.1098/rstl.1735.0014.

Hammarfelt, B., Rushforth, A., & de Rijcke, S. (2020). Temporality in academic evaluation: 'Trajectoral thinking' in the assessment of biomedical researchers. *Valuation Studies, 7*, 33. doi: 10.3384/VS.2001-5992.2020.7.1.33.

Hasselmann, K. (1976). Stochastic climate models 1, theory. *Tellus, 28*(6), 473-485. doi: 10.1111/j.2153-3490.1976.tb00696.x.

Haunschild, R. (2019). Which are the most influential cited references in *Information*? *Information, 10*(12). doi: 10.3390/info10120395.

Haunschild, R. (2021). BibPlots: Plot functions for use in bibliometrics. R package version 0.0.8. Retrieved 10 February, 2022, from https://cran.r-project.org/web/packages/BibPlots/index.html

Haunschild, R., & Bornmann, L. (2021). Reference Publication Year Spectroscopy (RPYS) in practice: A software tutorial. arXiv:2109.00969. Retrieved from https://arxiv.org/abs/2109.00969

Haunschild, R., & Marx, W. (2020). Discovering seminal works with marker papers. *Scientometrics, 125*(3), 2955-2969. doi: 10.1007/s11192-020-03358-z.

Hays, J. D., Imbrie, J., & Shackleton, N. J. (1976). Variations in the earth's orbit: Pacemaker of the ice ages. *Science, 194*(4270), 1121-1132. doi: 10.1126/science.194.4270.1121.

Hemlin, S., Allwood, C. M., Martin, B., & Mumford, M. D. (2013). Introduction: Why is leadership important for creativity in science, technology, and innovation. In S. Hemlin, C. M. Allwood, B. Martin & M. D. Mumford (Eds.), *Creativity and leadership in science, technology, and innovation* (pp. 1-26). New York, NY, USA: Taylor & Francis.

Holton, G., Chang, H., & Jurkowitz, E. (1996). How a scientific discovery is made: A case history. *American Scientist, 84*(4), 364-375.

Kokol, P., Blažun Vošner, H., & Završnik, J. (2021). Application of bibliometrics in medicine: A historical bibliometrics analysis. *Health Information & Libraries Journal, 38*(2), 125-138. doi: 10.1111/hir.12295.

Kuhn, T. S. (1962). *The structure of scientific revolutions*. Chicago, IL, USA: University of Chicago Press.

Leydesdorff, L., Bornmann, L., Comins, J., Marx, W., & Thor, A. (2016). Referenced Publication Year Spectrography (RPYS) and algorithmic historiography: The bibliometric reconstruction of András Schumbert's Œuvre. In W. Glänzel & B. Schlemmer (Eds.), *András Schubert - a world of models and metrics. Festschrift for András Schubert's 70th birthday* (pp. 79-96): International Society for Scientometrics and Informetrics.

Li, X., Yao, Q., Tang, X. L., Li, Q., & Wu, M. J. (2020). How to investigate the historical roots and evolution of research fields in China? A case study on iMetrics using RootCite. *Scientometrics, 125*(2), 1253-1274. doi: 10.1007/s11192-020-03659-3.

Lorenz, E. N. (1963). Deterministic non-periodic flow. *Journal of the Atmospheric Sciences, 20*(2), 130–141.

Manabe, S. (1969). Climate and ocean circulation .I. Atmospheric circulation and hydrology of earths surface. *Monthly Weather Review, 97*(11), 739-774. doi: 10.1175/1520-0493(1969)097<0739:Catoc>2.3.Co;2.

Manabe, S., & Stouffer, R. J. (1979). $CO_2$-climate sensitivity study with a mathematical-model of the global climate. *Nature, 282*(5738), 491-493. doi: 10.1038/282491a0.

Manabe, S., & Stouffer, R. J. (1980). Sensitivity of a global climate model to an increase of $CO_2$ concentration in the atmosphere. *Journal of Geophysical Research-Oceans, 85*(Nc10), 5529-5554. doi: 10.1029/JC085iC10p05529.

Manabe, S., & Wetheral, R. T. (1967). Thermal equilibrium of atmosphere with a given distribution of relative humidity. *Journal of the Atmospheric Sciences, 24*(3), 241-259. doi: 10.1175/1520-0469(1967)024<0241:Teotaw>2.0.Co;2.

Manabe, S., & Wetherald, R. T. (1975). Effects of doubling $CO_2$ concentration on climate of a general circulation model. *Journal of the Atmospheric Sciences, 32*(1), 3-15. doi: 10.1175/1520-0469(1975)032<0003:Teodtc>2.0.Co;2.

Manabe, S., & Wetherald, R. T. (1980). Distribution of climate change resulting from an increase in $CO_2$ content of the atmosphere. *Journal of the Atmospheric Sciences, 37*(1), 99-118. doi: 10.1175/1520-0469(1980)037<0099:Otdocc>2.0.Co;2.

Marx, W. (2021). History of RPYS. Retrieved March 7, 2022, from https://doi.org/10.6084/m9.figshare.14910615.v1

Marx, W., Bornmann, L., & Barth, A. (2013). Detecting the historical roots of research fields by reference publication year spectroscopy (RPYS). In J. Gorraiz, E. Schiebel, C. Gumpenberger, M. Horlesberger & H. Moed (Eds.), *14th International Society of Scientometrics and Informetrics Conference (ISSI)* (pp. 493-506). Vienne, Austria: University of Vienna.

Marx, W., Bornmann, L., Barth, A., & Leydesdorff, L. (2014). Detecting the historical roots of research fields by reference publication year spectroscopy (RPYS). *Journal of the Association for Information Science and Technology, 65*(4), 751-764. doi: 10.1002/asi.23089.

Marx, W., Haunschild, R., & Bornmann, L. (2017). Global warming and tea production: The bibliometric view on a newly emerging research topic. *Climate, 5*(3). doi: 10.3390/cli5030046.

Marx, W., Haunschild, R., Thor, A., & Bornmann, L. (2017). Which early works are cited most frequently in climate change research literature? A bibliometric approach based on Reference Publication Year Spectroscopy. *Scientometrics, 110*(1), 335-353. doi: 10.1007/s11192-016-2177-x.

Millán, J. D., Cudina, J. N., Ossa, J. C., Vega-Arce, M., Scholten, H., & Salas, G. (2020). Academic networks of critical social psychology in Brazil. An analysis of the impact and the intellectual roots. *Current Psychology*. doi: 10.1007/s12144-020-00827-9.

National Academy of Sciences. (2014). *Biographical memories: Stephen Schneider (1945-2010)*. Washington DC, CO, USA: National Academy of Sciences.

Newell, R. E., & Dopplick, T. G. (1979). Questions concerning the possible influence of anthropogenic $CO_2$ on atmospheric-temperature. *Journal of Applied Meteorology, 18*(6), 822-825. doi: 10.1175/1520-0450(1979)018<0822:Qctpio>2.0.Co;2.

NobelPrize.org. (2021). The Nobel Prize in physics 2021. Retrieved February 25, 2022, from https://www.nobelprize.org/prizes/physics/2021/summary

Plass, G. N. (1956). The carbon dioxide theory of climatic change. *Tellus, 8*(2), 140-154.

Ponomarev, I. V., Williams, D. E., Hackett, C. J., Schnell, J. D., & Haak, L. L. (2014). Predicting highly cited papers: A method for early detection of candidate breakthroughs. *Technological Forecasting and Social Change, 81*, 49-55. doi: 10.1016/j.techfore.2012.09.017.

R Core Team. (2018). R: A language and environment for statistical computing (Version 3.5.0). Vienna, Austria: R Foundation for Statistical Computing. Retrieved from https://www.r-project.org/

Ramanathan, V. (1976). Radiative transfer within the earth's troposphere and stratosphere: A simplified radiative-convective model. *Journal of Atmospheric Sciences, 33*(7), 1330-1346. doi: 10.1175/1520-0469(1976)033<1330:Rtwtet>2.0.Co;2.

Ramanathan, V., Lian, M. S., & Cess, R. D. (1979). Increased atmospheric $CO_2$: Zonal and seasonal estimates of the effect on the radiation energy-balance and surface-temperature. *Journal of Geophysical Research-Oceans, 84*(Nc8), 4949-4958. doi: 10.1029/JC084iC08p04949.

Rasool, S. I., & Schneider, S. H. (1971). Atmospheric carbon dioxide and aerosols: Effects of large increases on global climate. *Science, 173*(3992), 138-141. doi: 10.1126/science.173.3992.138.

Redner, S. (2005). Citation statistics from 110 years of *Physical Review*. *Physics Today, 58*(6), 49-54. doi: 10.1063/1.1996475.

Revelle, R., & Suess, H. E. (1957). Carbon dioxide exchange between atmosphere and ocean and the question of an increase of atmospheric $CO_2$ during the past decades. *Tellus, 9*(1), 18-27. doi: 10.1111/j.2153-3490.1957.tb01849.x.

Schneider, J. W., & Costas, R. (2017). Identifying potential "breakthrough" publications using refined citation analyses: Three related explorative approaches. *Journal of the Association for Information Science and Technology, 68*(3), 709-723.

Schumpeter, J. A. (1939). *Business cycles: A theoretical, historical and statistical analysis of the capitalist process*. London, UK: McGraw-Hill Book.

Sellers, W. D. (1969). A global climatic model based on the energy balance of the earth-atmosphere system. *Journal of Applied Meteorology and Climatology, 8*(3), 392-400. doi: 10.1175/1520-0450(1969)008<0392:Agcmbo>2.0.Co;2.

Shibayama, S., & Wang, J. (2020). Measuring originality in science. *Scientometrics, 122*, 409–427. doi: 10.1007/s11192-019-03263-0.

Small, H. (1973). Co-citation in the scientific literature: A new measure of the relationship between two documents. *Journal of the American Society for Information Science, 24*(4), 265-269. doi: 10.1002/asi.4630240406.

Stommel, H. (1961). Thermohaline convection with 2 stable regimes of flow. *Tellus, 13*(2), 224-230. doi: 10.1111/j.2153-3490.1961.tb00079.x.

Tyndall, J. (1861). On the absorption and radiation of heat by gasses and vapours, and on the physical connection of radiation, absorption, and conduction. *Philosophical Magazine, 4*(22), 273-285.

Vanderkam, D., Allaire, J. J., Owen, J., Gromer, D., & Thieurmel, B. (2018). dygraphs: Interface to 'Dygraphs' interactive time series charting library. R package version 1.1.1.6. Retrieved August 19, 2021, from https://CRAN.R-project.org/package=dygraphs

Vosner, H. B., Carter-Templeton, H., Zavrsnik, J., & Kokol, P. (2020). Nursing informatics: A historical bibliometric analysis. *Computers Informatics Nursing, 38*(7), 331-337. doi: 10.1097/cin.0000000000000624.

Wagner, C. S., Whetsell, T. A., & Mukherjee, S. (2019). International research collaboration: Novelty, conventionality, and atypicality in knowledge recombination. *Research Policy, 48*(5), 1260-1270. doi: 10.1016/j.respol.2019.01.002.

Walker, G. T., & Bliss, E. W. (1932). World weather V. *Memoirs of the Royal Meteorological Society, 4*(36), 53–84.

Wang, W. C., Yung, Y. L., Lacis, A. A., Mo, T., & Hansen, J. E. (1976). Greenhouse effects due to man-made perturbations of trace gases. *Science, 194*(4266), 685-690. doi: 10.1126/science.194.4266.685.

Winnink, J. J., Tijssen, R. J. W., & van Raan, A. F. J. (2018). Searching for new breakthroughs in science: How effective are computerised detection algorithms? *Technological Forecasting and Social Change*. doi: 10.1016/j.techfore.2018.05.018.

Yan, Y., Tian, S., & Zhang, J. (2020). The impact of a paper's new combinations and new components on its citation. *Scientometrics, 122*(2), 895-913. doi: 10.1007/s11192-019-03314-6.

Yao, Q., Li, X., Luo, F., Yang, L., Liu, C., & Sun, J. (2019). The historical roots and seminal research on health equity: A referenced publication year spectroscopy (RPYS) analysis. *International Journal for Equity in Health, 18*(1), 152. doi: 10.1186/s12939-019-1058-3.

Yeung, A. W. K. (2020). The historical roots of carotid artery stenting literature: An analysis of cited references. *Current Science, 119*(3), 447-450. doi: 10.18520/cs/v119/i3/447-450.

Yun, B., Lee, J. Y., & Ahn, S. (2020). The intellectual structure of women's studies: A bibliometric study of its research topics and influential publications. *Asian Women, 36*, 1-23. doi: 10.14431/aw.2020.6.36.2.1.

Zhao, Y., Han, J., Du, J., & Wu, Y. (2020). Origin and impact: A study of the intellectual transfer of Professor Henk F. Moed's works by using Reference Publication Year Spectroscopy (RPYS). In C. Daraio & W. Glänzel (Eds.), *Evaluative informetrics: The art of metrics-based research assessment. Festschrift in honour of Henk F. Moed* (pp. 145-162). Cham, Switzerland: Springer International Publishing.